\documentclass[aps,prl,twocolumn,unsortedaddress,showpacs]{revtex4}
\usepackage{graphicx}

\begin{document}

\title{High Frequency dynamics in metallic glasses}

\author{T. Scopigno$^{1}$, J.-B. Suck$^{2}$, R. Angelini$^{1}$, F. Albergamo$^{3}$, G.
Ruocco$^{1}$}

\affiliation{ $^{1}$INFM CRS-SOFT and Dipartimento di Fisica,
Universit\'a di Roma "La Sapienza",~I-00185, Roma, Italy
\email{tullio.scopigno@phys.uniroma1.it,
suck@physik.tu-chemnitz.de}\\
$^{2}$Institute of Physics, University of Technology Chemnitz, D-09107 Chemnitz, Germany\\
$^{3}$European Synchrotron Radiation Facility, F-38043 Grenoble,
Cedex , France}

\date{\today}

\begin{abstract}

Using Inelastic X-ray Scattering we studied the collective
dynamics of the glassy alloy Ni$_{33}$Zr$_{67}$ in the first
pseudo Brillouin zone, an energy-momentum region still unexplored
in metallic glasses. We determine key properties such as the
momentum transfer dependence of the sound velocity and of the
acoustic damping, discussing the results in the general context of
recently proposed pictures for acoustic dynamics in glasses.
Specifically, we demonstrate the existence in this strong glass of
well defined (in the Ioffe Regel sense) acoustic-like excitations
well above the Boson Peak energy.

\end{abstract}


\maketitle

The physics of disordered materials has attracted considerable
interest during recent years \cite{111}. This is mainly due to new
theoretical concepts that can handle systems lacking long-range
order, to large scale computer simulations based on realistic
interactions and to the simultaneous development of new
experimental techniques. Disordered materials show several kinds
of universal behaviors that differ from those of the ordered
state. In particular, great interest has been devoted to their
high frequency (THz) dynamics in respect to the corresponding
crystalline counterpart. In this context, Inelastic Neutron
Scattering (INS) investigations of the collective dynamics in
simple metallic glasses, which may be regarded as model systems,
have played a leading role \cite{suck83,hafn83,suck92,benm99}.
Because of their comparably simple atomic structure and their
nearly isotropic interaction, metallic glasses (and glasses with
Lennard-Jones (L-J) interaction), were the first tackled in
extended computer simulations, and one-to-one comparisons with
experiments were often undertaken. Simple systems like
Mg$_{70}$Zn$_{30}$ belong however to the minority of metallic
glasses and interest has therefore moved soon to more abundant and
characteristic representatives like the Zr-based transition
metal-transition metal (TM-TM) or the TM-metalloid glasses, which
are characterized by a strong topological and chemical short range
order \cite{hafn94}.

An important role in this context plays the TM-TM glass NiZr,
which can be produced by melt spinning in a large range of
compositions. Of these, the Zr-rich concentrations, for which also
crystalline NiZr$_{2}$ exists, have played a prominent role in the
investigation of the dynamics of metallic glasses
\cite{hafn94,aiha95}. Concerning collective dynamics, in
particular, the dispersion of modes around $Q_p$, (the $Q$
position of the main peak of the static structure factor), i.e. in
the region of maximal diffuse umklapp scattering \cite{hafn81},
was measured first for glassy Ni$_{24}$Zr$_{76}$ \cite{suck85} and
then in Ni$_{67}$Zr$_{33}$ \cite{otom98}. In these pioneering
experiments it was possible to prove the existence of Pseudo
Brillouin Zones with clear Zone Boundaries near the maxima of the
static structure factor, these latter acting like ``smeared out
lattice points'' \cite{hafn83}. However, important questions
concerning the properties of the collective excitations remained
unanswered on the experimental side: the existence of a low $Q$
linear dispersion, the wave-vector and energy transfer dependence
of the velocity and the damping of the excitations. In this
context, of importance here, is the recently proposed existence of
a crossover frequency, which should mark the transition between
different dynamical regimes \cite{ruff_prl,rufflille}.
This crossover frequency could be identified in different ways. A
first possibility ($\Omega_{IR}$) is by the Ioffe-Regel criterion,
i.e. by the condition $\Omega / \pi = \Gamma$, $\Omega$ being the
excitation frequency and $\Gamma$ the sound attenuation.
Furthermore, a crossover ($\Omega_{co}$) could be associated to
the frequency where the acoustic damping $\Gamma$ (i.e. the
broadening of the measured excitations) crosses from a $\Gamma
\propto \omega^4$ behavior ($\omega<\Omega_{co}$) to a weaker
power law, $\Gamma \propto \omega^{\alpha}$ with $\alpha \simeq 2$
($\omega>\Omega_{co}$). Finally, the crossover frequency could be
identified with the frequency of the Boson Peak (BP), i.e. the
maximum in the excess of vibrational density of states $g(\omega)$
with respect to the Debye behavior of the corresponding crystal,
represented as $\frac{g(\omega)}{\omega^2}$. Recently, the
coincidence of these three different definition of crossover
frequencies has been proposed to hold, at least for the case of
strong glasses \cite{ruff_prl,rufflille}.

In this letter we report the determination of the collective
dynamics in a metallic glass (Ni$_{33}$Zr$_{67}$), focused to the
region below $Q_p$, where the excitations are still very well
defined, aiming to answer some of the questions on the nature of
these high frequency modes, unanswered up to now  by INS
investigations. Using Inelastic -X-ray Scattering (IXS), we
exploited the lack of any kinematic restrictions in the accessible
$Q$-$E$ region which, combined with the good resolution, allowed
us to investigate previously inaccessible dynamics in metallic
glasses. A well defined acoustic branch has been observed up to $Q
\simeq Q_p$, and the Ioffe-Regel criterium indicates that
$\Omega_{IR}$ is close to $\Omega (Q_p)$, i.e. well above the
Boson Peak(BP) frequency. The behavior of the sound attenuation
can be rationalized in terms of a relaxation process related to
the structural disorder, similarly to what was found in simple
monatomic systems \cite{gcr_prlsim,scop_presim}. No evidence for a
Rayleigh regime ($\Gamma \propto \omega^4$) could be observed in
the explored $Q$ range, thus concluding that $\hbar \Omega{co}$,
if existing, is well below 5 meV. Finally, the fragility value
($m=26$), as deduced by the temperature dependence of the non
ergodicity parameter, agrees well with independent viscosity
determinations \cite{gue_phd}, thus corroborating the recent
proposed correlation between the low temperature vibrational
regime and the high temperature diffusive dynamics in glass
formers \cite{sco_sci}, and extending it to a class of systems not
considered so far. The whole scenario reported here does not
indicate the existence in this system of a single crossover
frequency marking the BP position ($\hbar \Omega_{BP}\simeq$ 3
meV), the Ioffe-Regel criterium ($\hbar \Omega_{IR}\simeq$ 8 meV)
and the end of Rayleigh scattering of phonons regime (which, if it
exists, has to be confined at energies $\hbar \Omega{co}< 5$ meV).

The experiment was performed at room temperature ($T$=297 K) at
the IXS beam-line ID16 of ESRF \cite{ver_strum,mas_strum} at fixed
scattering angle (and therefore fixed Q-values) in a region
between 1.5 and 20 nm$^{-1}$. The Q-resolution (FWHM) was set to
$\delta Q \approx 0.35$ nm$^{-1}$ and improved to $\delta Q
\approx 0.1$ nm$^{-1}$ at the lowest $Q$'s. Using the (9 9 9)
reflection for the Si monochromator and crystal analyzers the
overall energy resolution (FWHM) was $\delta \hbar\omega = 3.0$
meV. A five-analyzers bench, operating in horizontal scattering
geometry, allowed the simultaneous collection of spectra at five
different values of $Q$. Each energy scan from $-50 < \hbar\omega
< 50$ meV, where $\hbar\omega$ is the energy transfer $(E_0-E)$,
with $E_0$ and $E$ being the energy of the incident (17794 eV) and
the scattered X-ray, took approximately 300 minutes, and several
scans have been summed up to improve the statistical accuracy. The
Ni$_{33}$Zr$_{67}$ sample was freshly prepared by melt spinning
techniques, its structure was checked by X-ray diffraction before
(and during) the measurement.

\begin{figure} [h]
\centering 
\includegraphics[width=.45\textwidth]{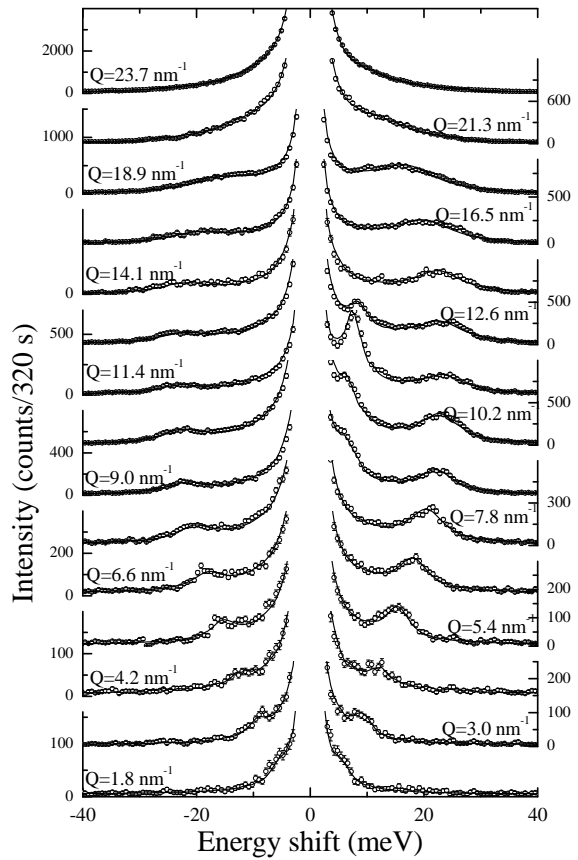}
\vspace{-.01cm} \caption{IXS spectra of Ni$_{33}$Zr$_{67}$ at the
indicated fixed $Q$ values (open dots). Also reported are the
best-fit line-shapes (continuous lines, see text). The low energy
structures at $E \approx 8$ meV observed around $Q\approx 11$
nm$^{-1}$ are spurious scattering from the kapton windows of the
sample container.} \label{panel}
\end{figure}

In Fig.\ref{panel} we report all the measured spectra in absolute
counts, after proper normalization for the incident monitor. The
elastic line is an effect of the non ergodicity typical of the
glassy state (scattering from static disorder) and possibly to
incoherent scattering from concentration fluctuations. Most
reasonably the measured inelastic signal can be ascribed to a
single acoustic propagating mode. The low energy ($E \approx $ 8
meV) peak observed around $Q\approx 11$ nm$^{-1}$ in the energy
loss side is due to the elastic scattering from the kapton window
located in the incoming beam before the sample. This peak, which
is due to the sample environment geometry and is only present in
the range $9 < Q < 13$ nm$^{-1}$, is also present in the empty
cell measurements and totally disappears when the empty cell is
subtracted from the data. The fitting procedure is performed by
adding the empty cell  with the proper transmission normalization
factor to the model line-shape used to represent the data.
Following generalized hydrodynamics, the simplest model function
to get the basic features of collective glassy dynamics is the
Damped Harmonic Oscillator (DHO), based on the  assumption of an
instantaneously decaying memory function for vibrational dynamics,
typical of Markovian processes, plus an elastic contribution due
to the arrested diffusive dynamics and to incoherent scattering
\cite{scop_rmp}. We have checked that fitting the data with more
refined model (Debye-like memory functions), while improving the
quality of the overall fit, does not produce significant variation
in the peak width and position. Therefore, for simplicity, we use
the DHO model function:

\begin{equation}
\frac{S(Q,\omega)}{S(Q)}=\left [ A(Q) \delta(\omega) +
 \frac{1-A(Q)}{\pi}\frac{\Omega^2(Q)\Gamma(Q)}
{(\omega^2-\Omega^2(Q))^2+\omega^2\Gamma^2(Q)} \right ]
\label{DHO}
\end{equation}

This function needs to be adapted to satisfy the detailed balance
condition and to be convoluted with the instrumental energy
resolution. The characteristic frequency of the acoustic mode,
$\Omega$, corresponds to  the maximum of the longitudinal current
spectra $C_L(Q,\omega) = (\omega/Q)^2 S(Q,\omega)$. Its $Q$
dependency defines the dispersion relation of the mode. The
parameter $\Gamma(Q)$ is related to the sound attenuation, while
$A(Q)$ is the intensity of the elastic scattering relative to the
total intensity of the spectrum at the Q under consideration.

\begin{figure} [h]
\centering
\includegraphics[width=.45\textwidth]{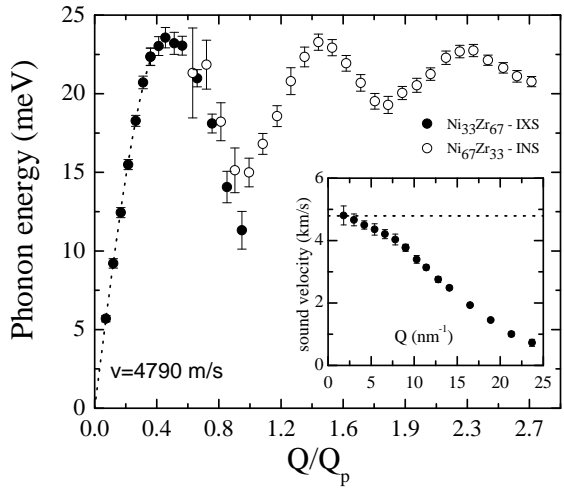}
\vspace{-4.5cm}\caption{Dispersion curve $\hbar\Omega(Q)$ of the
NiZr alloy as obtained from the IXS and INS experiments. The
momentum transfers have been scaled to $Q_p$ to account for the
different structural properties due to the different compositions
($Q_p^{(Ni_{33}Zr_{67})}=25.7$ nm$^{-1}$ and
$Q_p^{(Ni_{67}Zr_{33})}=29$ nm$^{-1}$). Inset: Q-dependence of the
sound velocity $v = \Omega / Q$ as measured by IXS.}
\label{svelocity}
\end{figure}

Based on the form of the dispersion reported in Fig.
\ref{svelocity}, we ascribe the corresponding excitations to
longitudinal acoustic modes. In spite of the fact that we were
able to measure at rather low Q-values, where the longitudinal
acoustic and optic branch should be well separated and the
excitations still well defined, we did not observe longitudinal
optic modes within the energy range covered in our experiment.
This is most likely due to the fact that the intensity of the
optic-like modes is expected to vanish in the $Q\rightarrow 0$
limit. Moreover, the dynamic structure factor measured by the IXS
is heavily dominated by the density-density correlations, while
the concentration-concentration contribution is negligible:
approximating the form factors of Ni and Zr by their $Q\rightarrow
0$ values, i.e. their atomic numbers, one has
$I_{IXS}(Q,\omega)\propto 2.25S_{\rho \rho}(Q,\omega)+
0.25S_{cc}(Q,\omega)- 1.5S_{\rho c}(Q,\omega)$. In the same
figure, we report the energy of the high frequency mode previously
measured with INS in Ni$_{67}$Zr$_{33}$. Due to the different
concentrations and cross-sections of the two species in the
neutron and X-ray investigations, leading to different weighting
factors in the total dynamic structure factor measured in each of
the two experiments, some difference in the measured excitation is
observed. Nevertheless normalizing the momentum transfer scale to
$Q_p$, one obtains a good overlap.


The sound attenuation, reported in Fig. 3a, shows a power law
dependence, with a change of slope around $Q \approx 7$ nm$^{-1}$
($E \approx 4$ meV) being proportional to $Q^2$ below this value
(full line) and to a lower $Q$ dependence above (dashed line). In
the same figure we also report the peak position $\Omega(Q)$
divided by $\pi$ (small full dots and thin full line). The
intersection of this line with $\Gamma(Q)$ define the Ioffe-Regel
limit, which is found here at $Q_{IR}$=14 $\pm$2 nm$^{-1}$ or
$\hbar \Omega_{IR}$=7 $\pm$1 meV. The Boson peak energy, on the
contrary, is $\hbar \Omega_{BP}$=3 meV \cite{suck96,hafn94}. This
observation clearly indicates that in the present glass there
exist well defined (in the Ioffe-Regel sense) acoustic excitations
well above the Boson peak energy. Furthermore, no transition
towards the Rayleigh $Q^4$ dependence is observed in the
investigated $Q$ range, thus posing an upper limit to the
crossover energy, $\hbar \Omega{co} < 4.5 meV$. We can therefore
conclude that neither the Boson peak energy, nor the crossover
energy, coincide with the Ioffe-Regel energy.

\begin{figure} [h]
\centering 
\includegraphics[width=.45\textwidth]{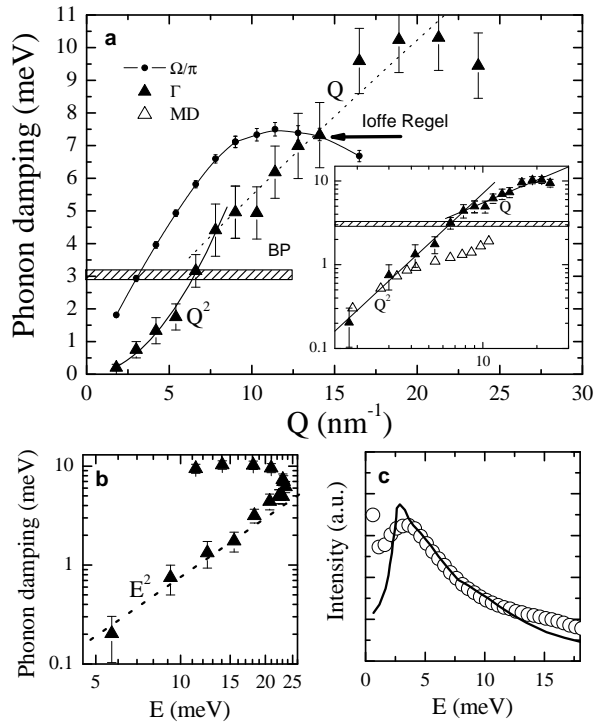}
\vspace{-1.5cm} \caption{(a): Q dependency of the damping of the
observed excitation (full triangles). A transition from the $Q^2$
to a linear dependence of the damping around $Q=7$ nm$^{-1}$ can
be observed. Values of $\Omega / \pi$ are also reported to
identify the Ioffe-Regel crossover (full circles), together with
the Boson Peak energy. Open triangles show the sound attenuation
calculated by means of molecular dynamics in a simple metal, which
shows a qualitatively similar behavior \cite{scop_presim}. (b):
Energy dependence of the damping: the breakdown of the quadratic
dependence is shifted to higher energies but not completely
removed. (c): reduced VDOS $\frac{g(\omega}{\omega ^2}$ for
Ni$_{33}$Zr$_{67}$ as measured by INS \cite{suck96} and calculated
by MD simulations \cite{hafn94}.} \label{atten}
\end{figure}

The $Q^2$ dependence is a well documented feature of the high
frequency attenuation in glasses, while the breakdown of this
dependence at intermediate $Q$ (around $Q$=7 nm$^{-1}$ here) is
not well understood up to now. In some systems, the breakdown of
the quadratic dependence is simply an effect of the modulation of
the structure of $S(Q)$, which can be wiped out reporting the
attenuation versus the excitation energy. In such cases, it seems
more significant to describe the sound attenuation as a $\Omega^2$
law \cite{scop_se}. This does not seem to be the case here, where
the $S(Q)$ is almost flat up to $10$ nm$^{-1}$, while the change
of slope is around 7 nm$^{-1}$, where the dispersion leaves its
linear slope. Indeed, in Fig. \ref{atten}b one can observe a
deviation from the $\Gamma \propto \Omega^2$ law above $\Omega
\approx 20$ meV ($Q \approx 8$ nm$^{-1}$). The observed transition
might be possibly related to a dynamical effect, as reported in
the molecular dynamics simulation of a simple metallic glass (open
triangles in fig \ref{atten}a), scaled by an arbitrary factor for
presentation reasons) \cite{scop_presim}.

\begin{figure} [h]
\centering
\includegraphics[width=.47\textwidth]{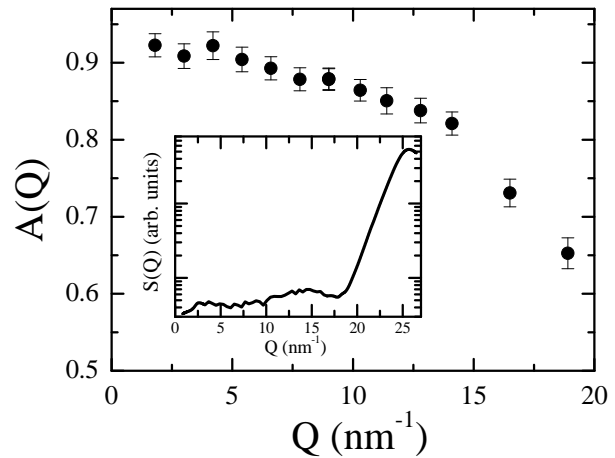}
\vspace{-5.1cm} \caption{Q-dependence of the elastic/total
scattered intensity. Inset: static structure factor, note the
prepeak at $Q \approx 14$ nm$^{-1}$), where A(Q) changes its
slope.} \label{nef}
\end{figure}

In Fig. \ref{nef}, the elastic/total intensity ratio
$A(Q)=\frac{I_{el}}{I_{tot}}$ is reported. Generally speaking, the
elastic signal comes from the relaxation processes active over a
timescale longer than the inverse of the phonon frequency. In
systems, where the dominant elastic contribution comes from the
structural relaxation, $A(Q)$ is a measure of the so called non
ergodicity factor, $f(Q)$. It has been recently proposed
\cite{sco_sci} that the low temperature behavior of the non
ergodicity factor in a $T/T_g$ plot is closely correlated to the
so called Angel plot, i.e. to the temperature behavior of
viscosity on approaching the glass transition in a $T_g/T$ scale.
More specifically, in terms of the fragility index $m = \lim_{T
\rightarrow T_g} \frac{d \log(\eta)}{d(T_g/T)}$ and of the
parameter $\alpha = \lim_{T \rightarrow 0}  \frac{d f(Q\rightarrow
0,T}{d(T/T_g)}$ one has $m\approx 135 \alpha$. At the same time,
the temperature dependence of the non ergodicity parameter in the
$T\rightarrow 0$ limit is well described by the functional form
$f(Q, T\rightarrow 0)=\frac{1}{1+\alpha T/T_g}$. The glass
transition temperature of Ni$_{33}$Zr$_{67}$ is about $T_g=652$ K,
and using our determination of $A(Q\rightarrow 0,T)\approx 0.92$
at room temperature one has $\alpha=0.19$, which would lead to a
fragility of about $m=26$, indicating Ni$_{33}$Zr$_{67}$ to
be a strong glass. Though this value of fragility has to be taken
as a lower estimate, due to possible elastic contribution from
concentration fluctuations, it is in very good agreement with
independent high temperature viscosity determination, which would
give $m=24$ \cite{gue_phd}.

Summing up, we have studied the high frequency dynamics of the
strong metallic glass Ni$_{33}$Zr$_{67}$ in a range of momentum
transfer below the first maximum of the static structure factor.
Evidence of a longitudinal acoustic-like branch has been reported,
which is the extension into the sensible low $Q$ region of the
high frequency excitations observed a few years ago by means of
INS. The analysis of the non-ergodicity factor allows us to
classify -in agreement with previous determination \cite{gue_phd}-
this glass as "strong" ($m$=26). At low $Q$, the sound attenuation
exhibits a $Q^2$ behavior, with no hint of a possible transition
towards a $Q^4$ dependence. Finally, we determine the Ioffe-Regel
limit, that is reached close to the top of the acoustic branch, at
$E_{IR}$=7 $\pm$1 meV, thus well above the Boson peak energy,
$E_{BP}$= 3 meV. In the present strong glass, there exist well
defined (in the Ioffe Regel sense) acoustic-like excitations well
above the Boson Peak energy. These findings are not in accordance
with recent models proposed for the acoustic properties in strong
glasses \cite{ruff_prl,rufflille} that predict $E_{IR}$=$E_{BP}$.

Acknowledgment: We would like to thank Mrs. Teichmann for the
melt spinning of the sample.


\end{document}